\documentclass{article}
\usepackage{spconf,amsmath,graphicx}
\usepackage{url}
\usepackage{multirow}
\usepackage{threeparttable}
\usepackage{xcolor,colortbl}
\usepackage{subfigure}

\usepackage{amssymb}
\usepackage{amsmath}
\usepackage{subfiles}
\usepackage{makecell}
\usepackage{listings} 
\usepackage{adjustbox}
\usepackage{arydshln}
\usepackage{cancel}
\usepackage{multibib}
\usepackage{hyperref}
\usepackage{threeparttable}
\usepackage{makecell}
\newcites{SI}{Supplementary References}

\newif\ifincludeSupplementaryOnly
\hypersetup{
    colorlinks=True,    
    pdfborder={0 0 0},   
}

\let\OLDthebibliography\thebibliography
\renewcommand\thebibliography[1]{
  \OLDthebibliography{#1}
  \setlength{\parskip}{0pt}
  \setlength{\itemsep}{0pt plus 0.3ex}
}

\begin{document}\sloppy

\def\x{{\mathbf x}}
\def\L{{\cal L}}
\ifincludeSupplementaryOnly
    \input{6_Supplementary.tex}
\else


\title{Low-Rank Adaptation of Pre-trained Vision Backbones for Energy-Efficient Image Coding for Machines}


%

\name{Yichi Zhang, Zhihao Duan, Yuning Huang, Fengqing Zhu}
\address{Elmore Family School of Electrical and Computer Engineering,\\Purdue University, West Lafayette, Indiana, U.S.A.\\\{zhan5096, duan90, huan1781, zhu0\}@purdue.edu}

\maketitle

\begin{abstract}
Image Coding for Machines (ICM) focuses on optimizing image compression for AI-driven analysis rather than human perception. Existing ICM frameworks often rely on separate codecs for specific tasks, leading to significant storage requirements, training overhead, and computational complexity. To address these challenges, we propose an energy-efficient framework that leverages pre-trained vision backbones to extract robust and versatile latent representations suitable for multiple tasks. We introduce a task-specific low-rank adaptation mechanism, which refines the pre-trained features to be both compressible and tailored to downstream applications. This design minimizes trainable parameters and reduces energy costs for multi-task scenarios. By jointly optimizing task performance and entropy minimization, our method enables efficient adaptation to diverse tasks and datasets without full fine-tuning, achieving high coding efficiency. Extensive experiments demonstrate that our framework significantly outperforms traditional codecs and pre-processors, offering an energy-efficient and effective solution for ICM applications. The code and the supplementary materials will be available at: {\url{https://gitlab.com/viper-purdue/efficient-compression}}.
\end{abstract}


\section{Introduction}
\label{sec:intro}

Recent learning-based image compression (LIC) methods have gained significant attention due to their simple framework and impressive performance. Several approaches~\cite{he2022elic,liu2023learned,zhang2024another,zhang2024theoretical} have even surpassed the most powerful rule-based method, VVC/H.266~\cite{bross2021overview} intra, in terms of rate-distortion (RD) performance. However, these codecs are designed primarily to optimize human visual perception. When applied to AI-driven analysis tasks, the effectiveness of existing image coding methods, including both rule-based codecs and LICs, remains debatable. The fundamental differences between the information requirements of intelligent tasks and human vision, coupled with the diversity and potential unknowns of downstream tasks, make existing codecs suboptimal for compressing images tailored to machine analysis.

To address this gap, a new research direction, \textit{image coding for machines (ICM)}~\cite{duan2020video,feng2023prompt,feng2022image,chen2023transtic,liu2023icmh,li2025image}, has emerged. ICM seeks to establish a unified framework that efficiently compresses images to support intelligent analytics for a wide range of applications. Existing ICM approaches can be categorized into three main branches~\cite{feng2023prompt}: (a) {Task-Specific Tuned Codecs:} These methods compress images and reconstruct them for subsequent intelligent analysis~\cite{li2025image,lu2024preprocessing}. (b) {Feature Compression:} These approaches extract and compress task-specific intermediate features for individual tasks~\cite{chen2019toward,yang2024video,sheng2024vnvc,tian2024coding}. Despite their advancements, both approaches have significant limitations: the need for separate codecs for different tasks, leading to poor generalization and substantial overhead in training computation and model storage. (c) {Generic Feature Compression:} This approach uses a universal feature extractor and codec to compress general features applicable to all tasks~\cite{feng2022image}. While this strategy improves generalization, it often sacrifices efficiency by overlooking task-specific characteristics during compression and downstream analysis. 

In this paper, we propose a new method for image coding for machines that addresses the shortcomings of existing approaches. Drawing inspiration from the success of parameter-efficient fine-tuning techniques such as~\cite{ding2023parameter,hu2022lora,liu2024dora,he2023parameter,agiza2024mtlora} and Prompt-ICM~\cite{feng2023prompt}, our method combines the strengths of task-specific feature compression and generic feature extraction. We introduce two core designs: 1. {Pre-Trained Vision Backbones:} We leverage pre-trained vision backbones trained on large-scale datasets and optimized with carefully designed representation learning strategies~\cite{bao2022beit,radford2021learning} to extract free versatile and robust representations at no additional training cost. The main pre-trained backbone remains fixed for all tasks. 2. {Low-Rank Adaptation Mechanism:} We enhance the pre-trained backbone with DoRA (a low-rank adaptation module) layers~\cite{liu2024dora} that are specifically optimized for each task. By carefully selecting the rank and adapted modules, our approach enables task-specific optimization with minimal trainable parameters, significantly reducing training overhead, storage requirement, and energy consumption while maintaining superior performance.

Our contributions are summarized as follows:
\begin{itemize}
\item We propose a new method that leverages pre-trained vision backbones to extract robust and versatile latent representations for ICM at no additional training cost.
\item We introduce a low-rank adaptation strategy that refines pre-trained features, achieving task-specific optimization and compressibility while minimizing parameter overhead and energy consumption.
\item Extensive experiments demonstrate that our method outperforms traditional codecs and pre-processors while achieving superior parameter efficiency and coding performance compared with full fine-tuning.
\end{itemize}

\section{Proposed Method}
\subsection{Framework Overview}

\begin{figure*}[htbp] 
    \centering 
    \includegraphics[width=0.8\linewidth]{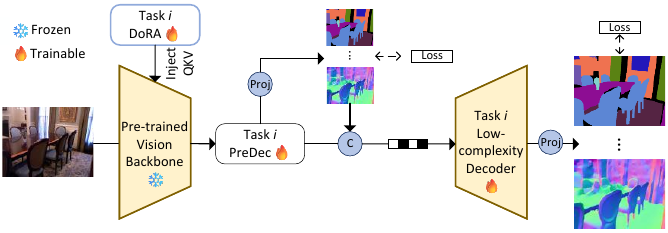}
    \caption{Overview of the proposed framework.  
    ``C'' denotes concatenation, and ``Proj'' refers to the linear projection layer.}
    \label{fig:overview} 
\end{figure*}

The overall framework of the proposed method is depicted in Fig.~\ref{fig:overview}. It comprises three components: (1) a pre-trained vision backbone encoder with task-specific DoRA layers, (2) task preliminary decoders, and (3) a low-complexity decoder.  

The pre-trained vision backbone encoder is shared across all tasks, extracting versatile and task-agnostic features from input images. To enable efficient task-specific optimization and compression, task-specific DoRA layers (low-rank adaptations) are introduced, providing tailored adaptation with minimal computational overhead. These adapted features are processed by task-specific preliminary decoders, which generate intermediate task-specific features and preliminary predictions for intermediate supervision. The preliminary predictions are upsampled by bicubic interpolation and optimized using task-specific labels and loss functions during training. Subsequently, the intermediate features and preliminary predictions are then concatenated along the channel dimension and compressed to form the decoder's input. Finally, a low-complexity transformer decoder is employed to model global spatial feature interactions, refining the task-specific features and preliminary predictions to produce the final output. This design ensures efficient task adaptation and high-quality results while maintaining low computational costs.

\subsection{Pretrained Vision Backbone with DoRAs}
To reduce training and storage costs, we employ a single general feature extractor with fixed parameters to extract versatile features for multiple downstream tasks. This eliminates the need to train separate encoders for individual tasks, leading to significant reductions in both training energy consumption and storage overhead. For robust generalization and effective representation, we adopt EVA-02~\cite{fang2024eva}, pre-trained on Merged-38M (IN-22K, CC12M, CC3M, COCO (train), ADE20K (train), Object365, and OpenImages) with masked image modeling (using EVA-CLIP as a MIM teacher) and fine-tuned on ImageNet-22k, as the general feature extractor.  

However, the general encoder alone does not account for task-specific characteristics or the compression process, which lead to suboptimal task performance. To address this, we introduce DoRA layers~\cite{liu2024dora} into the backbone. These layers efficiently adapt the versatile pre-trained features to achieve both task-specific optimization and feature compressibility. By integrating DoRAs, the backbone acts as a shared encoder, while each task benefits from its own lightweight, task-specific DoRA layers. This design ensures tailored adaptation for each task without the overhead of full fine-tuning.

\textbf{DoRA: Efficient Task-Specific Adaptation.}  As shown in Fig.~\ref{fig:dora}, DoRA~\cite{liu2024dora} decomposes the pre-trained weight \( W_0 \) into its magnitude and directional components, both of which are fine-tuned. LoRA~\cite{hu2022lora} is applied to the directional component \( V \) to decompose it into a low-rank form for fine-tuning.
\begin{figure}[htbp] 
\centering 
\includegraphics[width=1\linewidth]{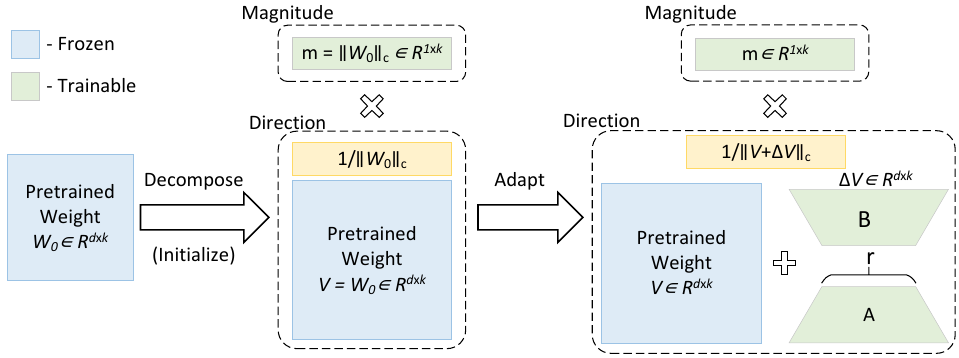} 
\caption{DoRA layer} 
\label{fig:dora} 
\end{figure}
Specifically, DoRA initializes the weight as:  
\(
W_0 = m \frac{V}{\|V\|_c} = \|W_0\|_c \frac{W_0}{\|W_0\|_c},
\) 
where \( m \in \mathbb{R}^{1 \times k} \) is the magnitude vector, \( V \in \mathbb{R}^{d \times k} \) is the directional matrix, and \( \| \cdot \|_c \) denotes the vector-wise norm across each column. This decomposition ensures that each column of \( \frac{V}{\|V\|_c} \) remains a unit vector, while \( m \) scales its magnitude. Here, \( V \) is frozen, and \( m \) is trainable. The directional component is updated by:  
\begin{equation}
W' = m \frac{V + \Delta V}{\|V + \Delta V\|_c} = m \frac{W_0 + BA}{\|W_0 + BA\|_c}.
\end{equation}
Here, \( \Delta V \) represents the low-rank directional update, where \( B \in \mathbb{R}^{d \times r} \) and \( A \in \mathbb{R}^{r \times k} \) are the low-rank matrices. These matrices are initialized such that \( W' \) equals \( W_0 \), ensuring a smooth initialization process. By utilizing a pre-trained vision backbone as a shared encoder and introducing lightweight DoRA layers for task-specific and compression adaptation, our method significantly reduces training costs compared to full fine-tuning or training from scratch. In terms of storage, our approach requires only a full copy of the pre-trained vision backbone and a few lightweight task-specific DoRA layers, which substantially lowers the storage requirements.

\subsection{Task-Specific Preliminary Decoders}
Inspired by previous work~\cite{ye2024invpt++}, the task-specific preliminary decoders are created to generate coarse preliminary predictions and task-specific features for each downstream task. Each decoder is built using a base unit consisting of a \( 3 \times 3 \) convolutional layer, followed by batch normalization and the GELU activation function, collectively denoted as {Conv-BN-GELU}. A task-specific preliminary decoder comprises two such {Conv-BN-GELU} units, which take the encoder features as input and produce refined task-specific features along with preliminary predictions. The preliminary predictions are supervised using ground-truth labels corresponding to each task via appropriate loss functions. The task-specific features and predictions are then concatenated along the channel dimension and compressed using a {dual spatial entropy model}~\cite{Wang2023evc}. The resulting compressed features serve as input to the subsequent transformer decoder.

\subsection{Low-Complexity Decoder}

On the receiver side, a low-complexity decoder is employed to produce the high-resolution final task output from the transmitted task-specific features and coarse preliminary predictions, such as segmentation masks for semantic segmentation. The decoder is designed as an efficient inverse version of the encoder with key revisions to reduce complexity while maintaining performance:

\textbf{Fewer Stages for Reduced Computation.}
The encoder consists of 4 stages, leading to a 16x spatial downsampling. To reduce computational complexity, the decoder includes only 3 stages, achieving 8x upsampling. The final 2x upsampling is performed using interpolation after the final projection. All the upsampling in the decoder is performed by bicubic interpolation.

\textbf{Low-Complexity Self-Attention.}  
Direct computation of self-attention on high-resolution task features \( \mathbf{F}_s \in \mathbb{R}^{H_s \times W_s \times C_s} \) (where \( H_s, W_s, C_s \) denote the spatial dimensions and channels of stage \( s \)) incurs significant memory overhead. To address this, we downsample the features before computing self-attention. For queries, a \( 3 \times 3 \) depth-wise convolution with a stride of 2 is applied, resulting in \( \mathbf{Q}_s = \mathbf{W}_s^q (\text{Conv}(\mathbf{F}_s)) \in \mathbb{R}^{\frac{H_s W_s}{4} \times C_s} \). For keys and values, a depth-wise convolution with a stride of \( k_s = 2^{s+1} \) is used, producing \( \mathbf{K}_s = \mathbf{W}_s^k (\text{Conv}(\mathbf{F}_s, k_s)) \) and \( \mathbf{V}_s = \mathbf{W}_s^v (\text{Conv}(\mathbf{F}_s, k_s)) \), where \( \mathbf{K}_s, \mathbf{V}_s \in \mathbb{R}^{\frac{H_s W_s}{k_s^2} \times C_s} \) and \( \mathbf{W}_s^q, \mathbf{W}_s^k, \mathbf{W}_s^v \) denote linear projection weights. The self-attention score matrix is then computed as \( \mathbf{A}_s = \mathbf{Q}_s \mathbf{K}_s^T \in \mathbb{R}^{\frac{H_s W_s}{4} \times \frac{H_s W_s}{k_s^2}} \). This design enables efficient self-attention at reduced resolutions. After performing the attention mechanism, the features are upsampled back to the original resolution via bicubic interpolation. The remaining operations follow standard EVA-02 configurations~\cite{fang2024eva}.

\textbf{Cross-scale self-attention.}
To incorporate multi-scale information, which is crucial for visual tasks~\cite{chen2021crossvit}, we introduce a fusion-based cross-scale self-attention. Specifically, the self-attention score matrix \( \mathbf{A}_{s-1} \) from the previous stage is upsampled via interpolation and added to the current stage’s score matrix \( \mathbf{A}_s \), scaled by a learnable parameter \( \alpha_s \):    $ \mathbf{A}_s \leftarrow \mathbf{A}_s + \alpha_s \cdot \text{Interp}(\mathbf{A}_{s-1})$. This mechanism effectively fuses multi-scale information, enhancing the overall performance of the transformer decoder. 

The outputs of the low-complexity decoder, which are the desired high-resolution task-specific outputs, are supervised using task-specific labels and corresponding loss functions.

\section{Experiments}
\subsection{Experimental Settings}

\textbf{Datasets.}
The experiments are conducted on two popular scene understanding datasets with multi-task labels, \textit{i.e.}, NYUD-v2, and PASCAL-Context. {NYUD-v2} supports semantic segmentation, monocular depth estimation, surface normal estimation, and object boundary detection. {PASCAL-Context} provides semantic segmentation, human parsing, object boundary detection, surface normal estimation, and saliency detection. We perform experiments on \textit{all} tasks.

\textbf{Training.} The model is trained using a rate-accuracy loss function, consistent with previous ICM approaches~\cite{feng2023prompt,feng2022image}: \(\mathcal{L} = R + \lambda \mathcal{L}_{\text{task}}\), where \(R\) represents the rate of transmitted features, and \(\mathcal{L}_{\text{task}}\) denotes the task-specific loss functions. The models are trained for 160,000 iterations, with a batch size of 2. Adam optimizer is adopted with a learning rate of 2e-5 and a weight decay rate of 1e-6. A polynomial learning rate scheduler is used. 

\textbf{Evaluation.}
Semantic segmentation (Semseg) and human parsing (Parsing) are evaluated with mean Intersection over Union (mIoU); monocular depth estimation (Depth) is evaluated with Root Mean Square Error (RMSE); surface normal estimation (Normal) is evaluated by the mean Error (mErr) of predicted angles; saliency detection (Saliency) is evaluated with maximal F-measure (maxF); object boundary detection (Boundary) is evaluated with the optimal-dataset-scale F-measure (odsF).

\begin{figure*}[htbp] 
\newcommand{\mywidth}{0.22}
\centering 
\subfigure[Semseg]{\includegraphics[width=\mywidth\linewidth]{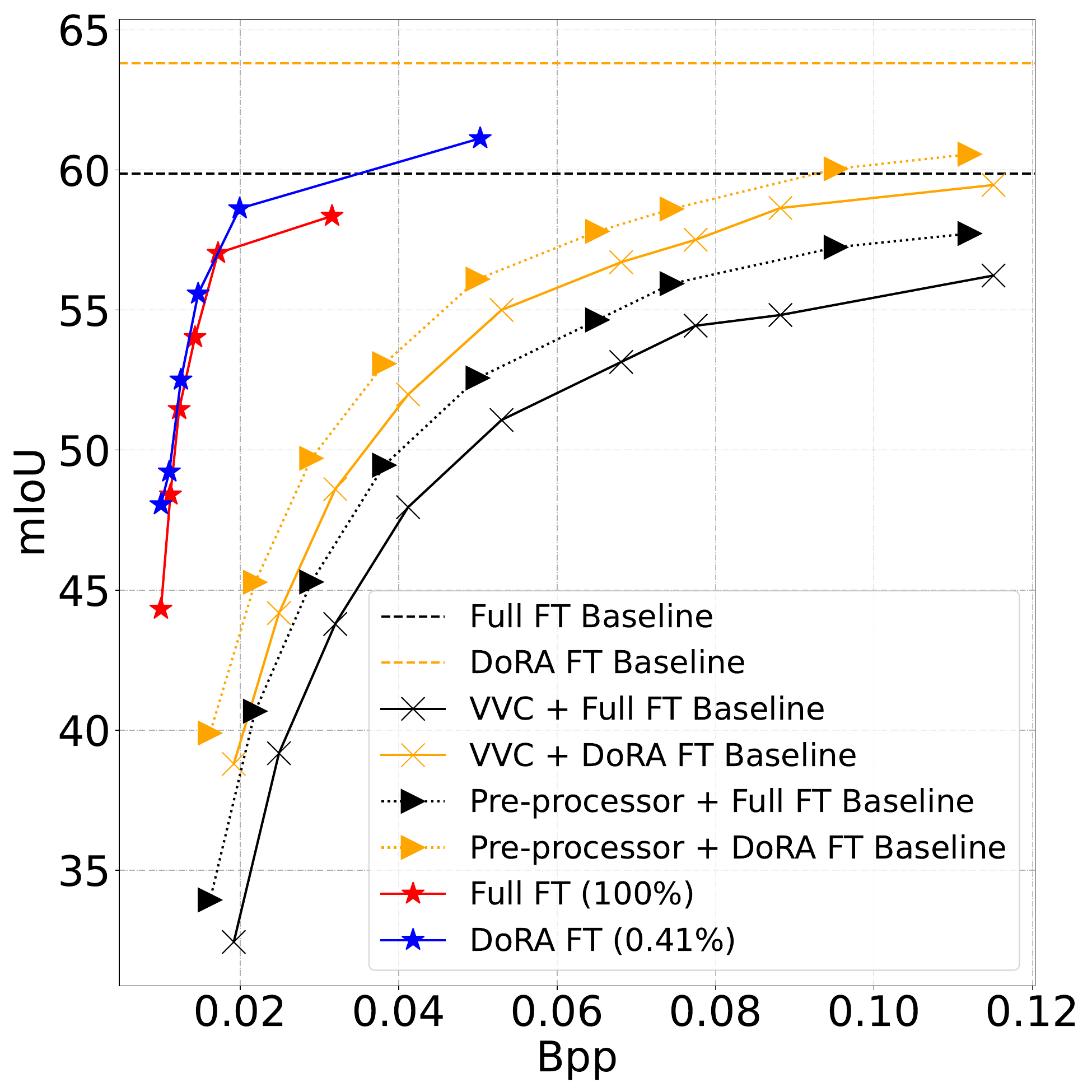}}
\subfigure[Depth]{\includegraphics[width=\mywidth\linewidth]{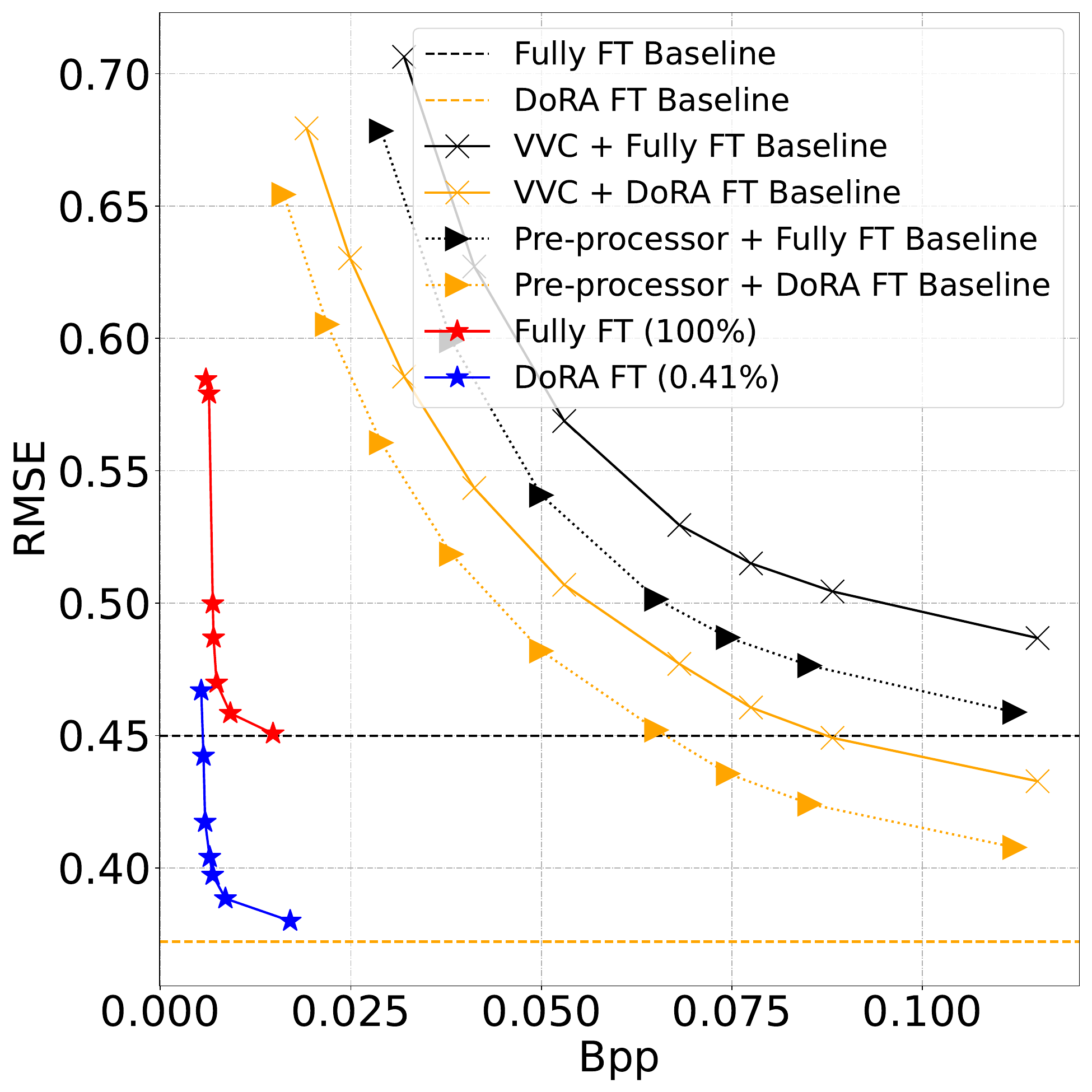}}
\subfigure[Normal]{\includegraphics[width=\mywidth\linewidth]{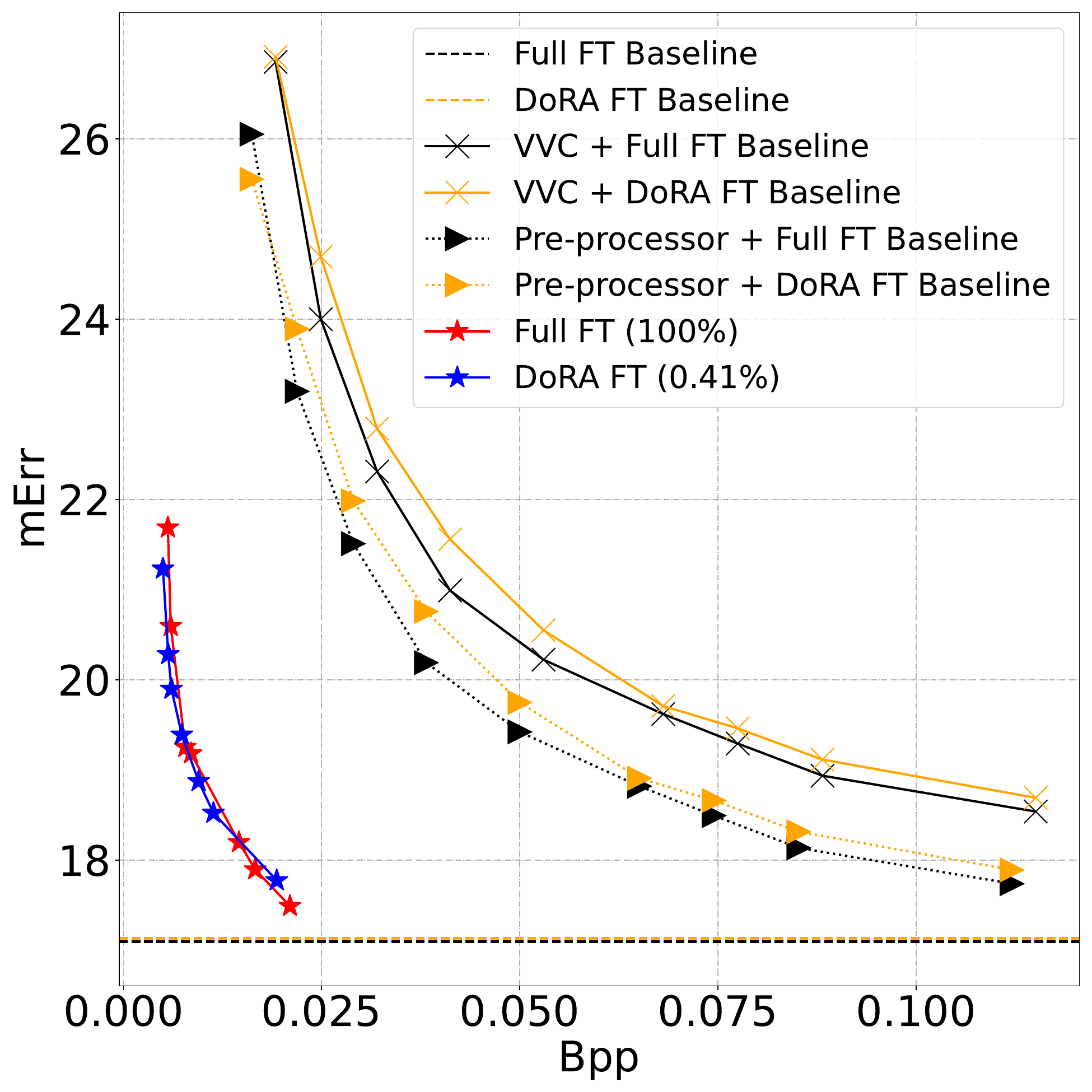}}
\subfigure[Boundary]{\includegraphics[width=\mywidth\linewidth]{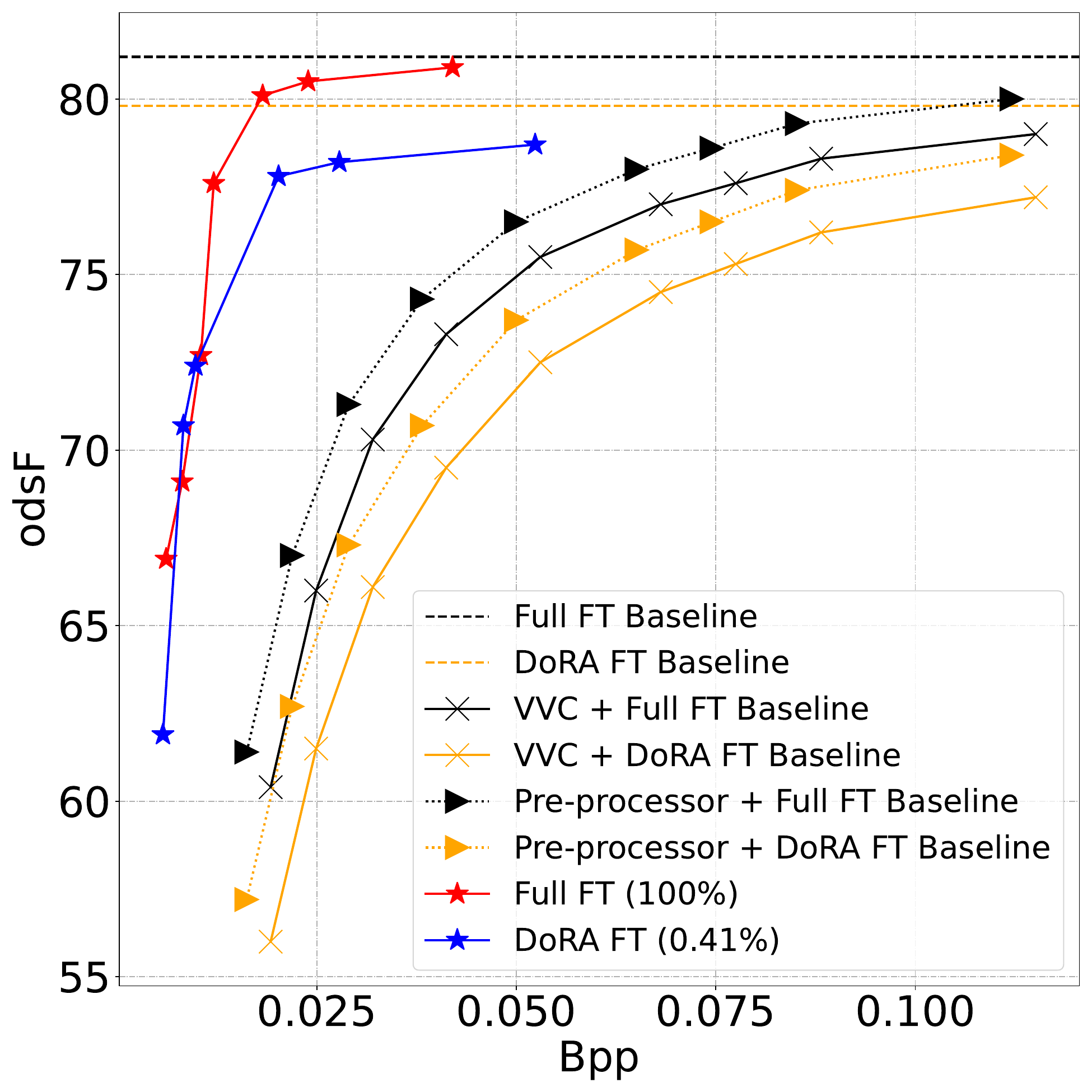}}
\caption{\textbf{RA curves of various methods on NYUD-v2 dataset. }{\it Please zoom in for more details}.} 
\label{fig:nyud_result} 
\end{figure*}

\begin{figure*}[htbp] 
\newcommand{\mywidth}{0.19}
\centering 
\subfigure[Semseg]{\includegraphics[width=\mywidth\linewidth]{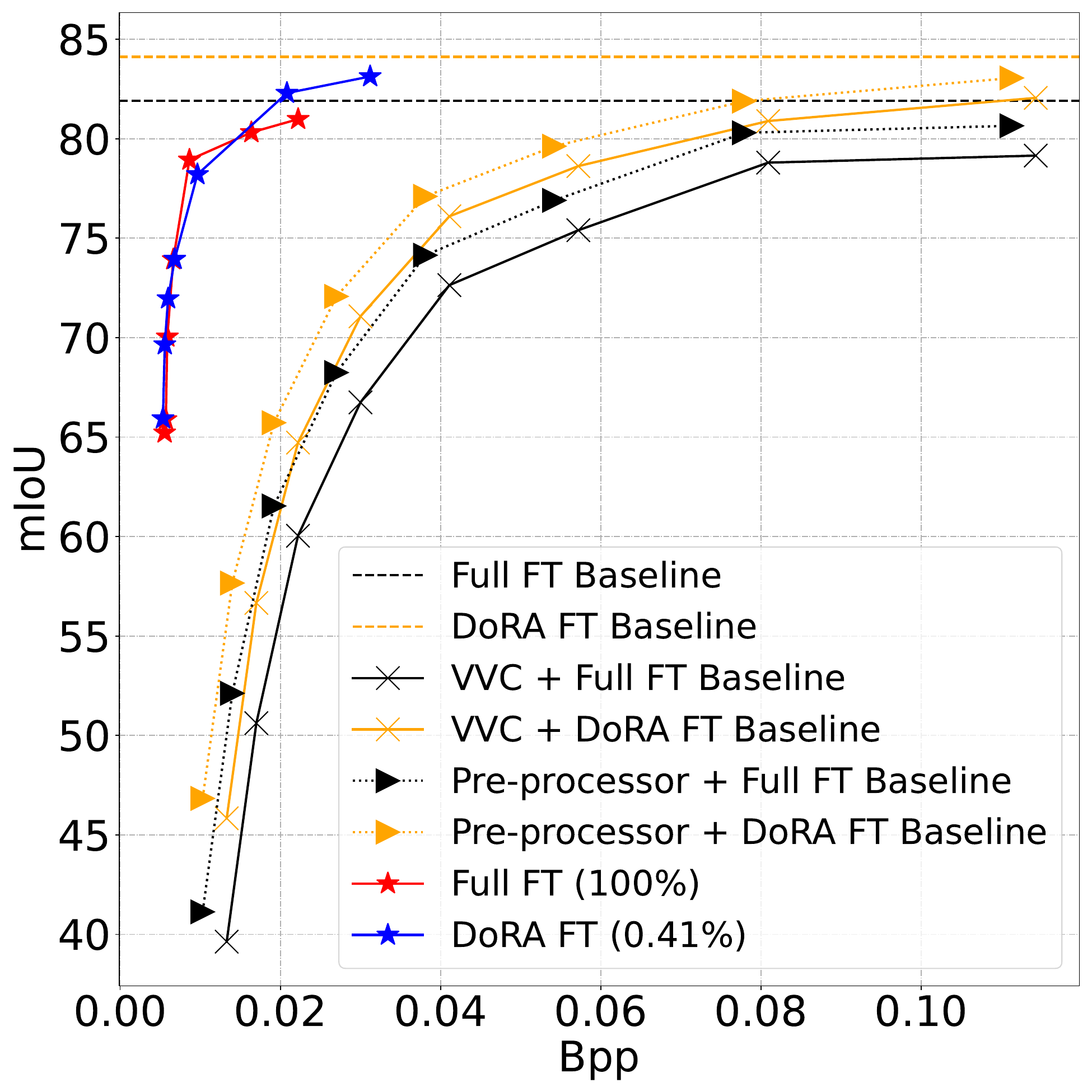}} 
\subfigure[Human]{\includegraphics[width=\mywidth\linewidth]{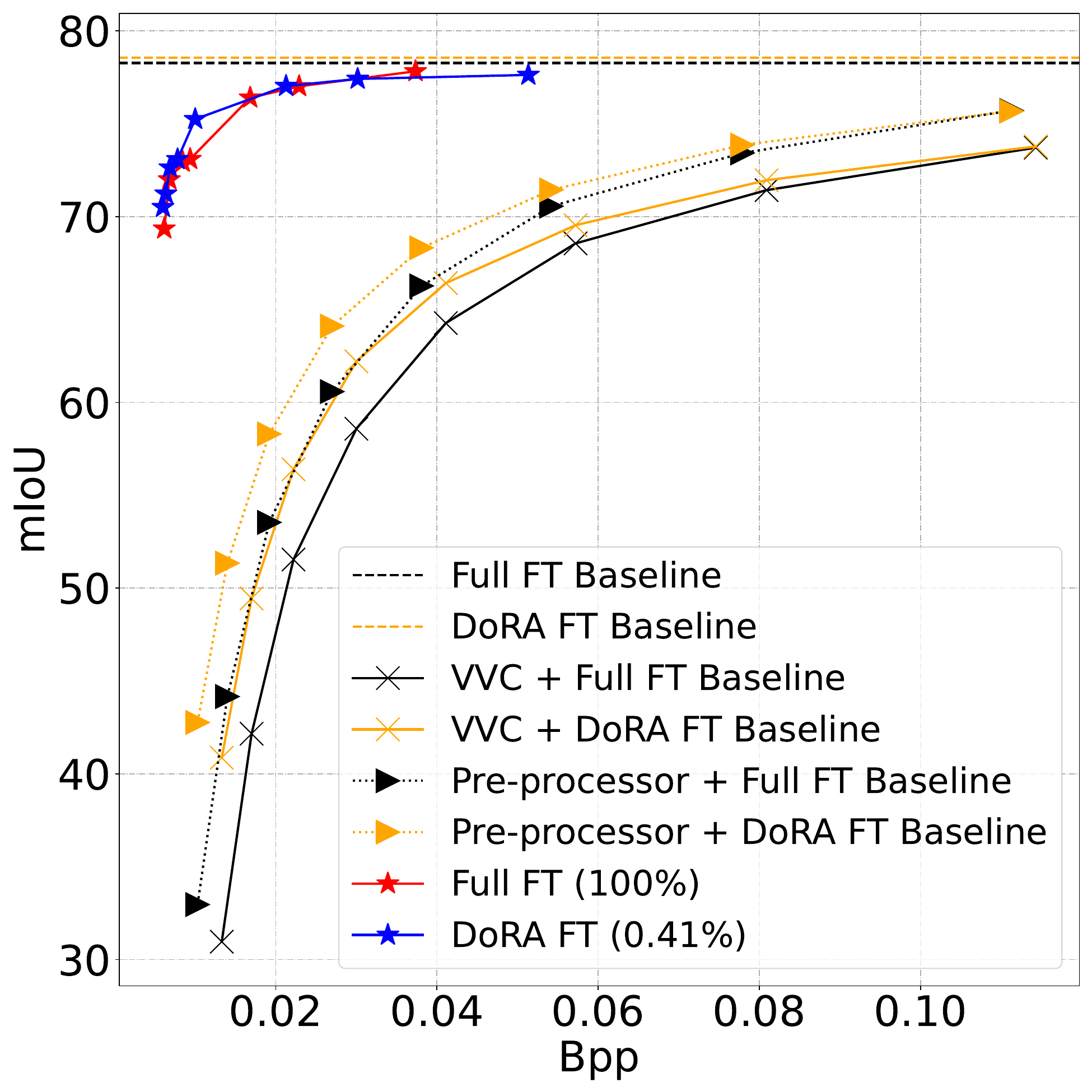}} 
\subfigure[Sal]{\includegraphics[width=\mywidth\linewidth]{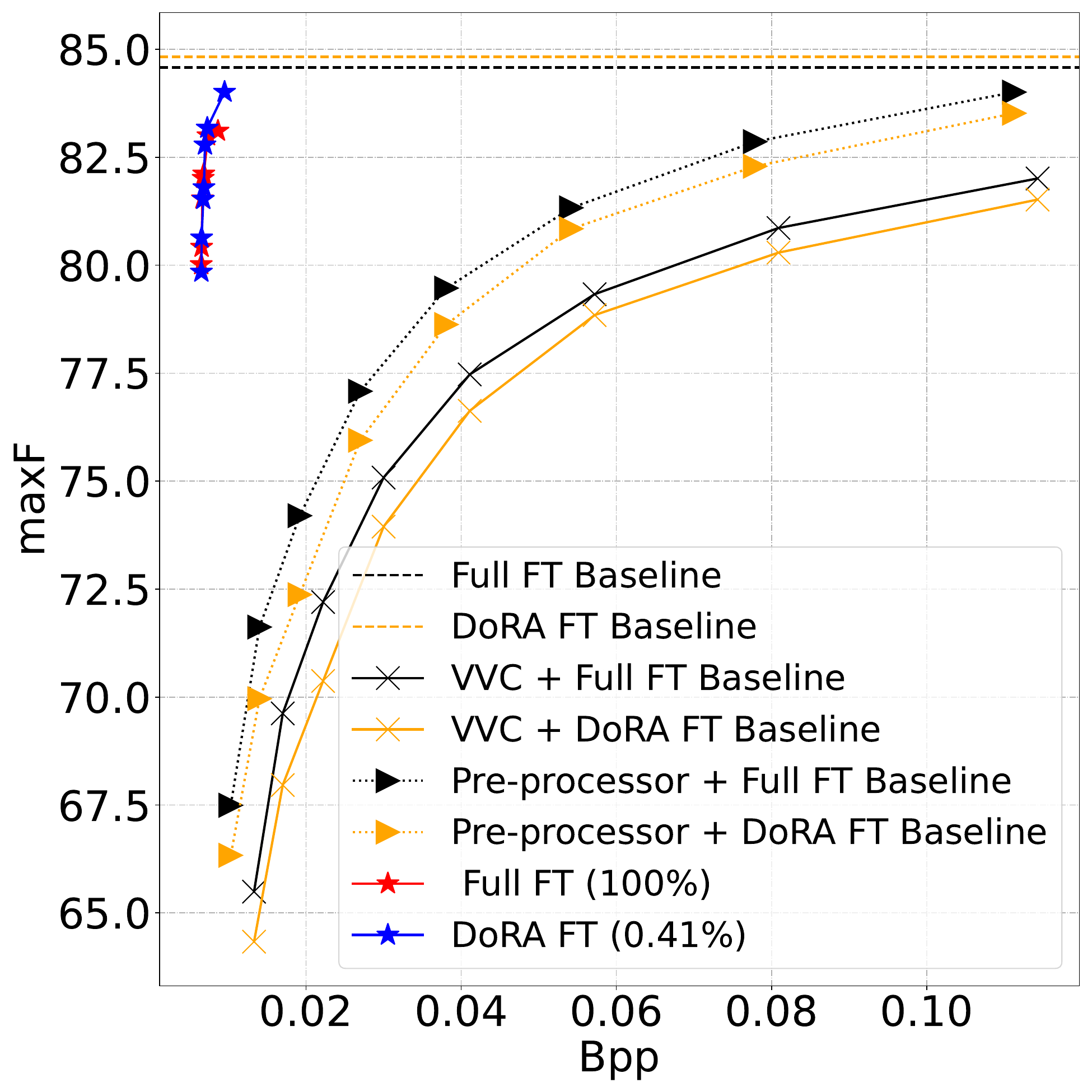}}
\subfigure[Normal]{\includegraphics[width=\mywidth\linewidth]{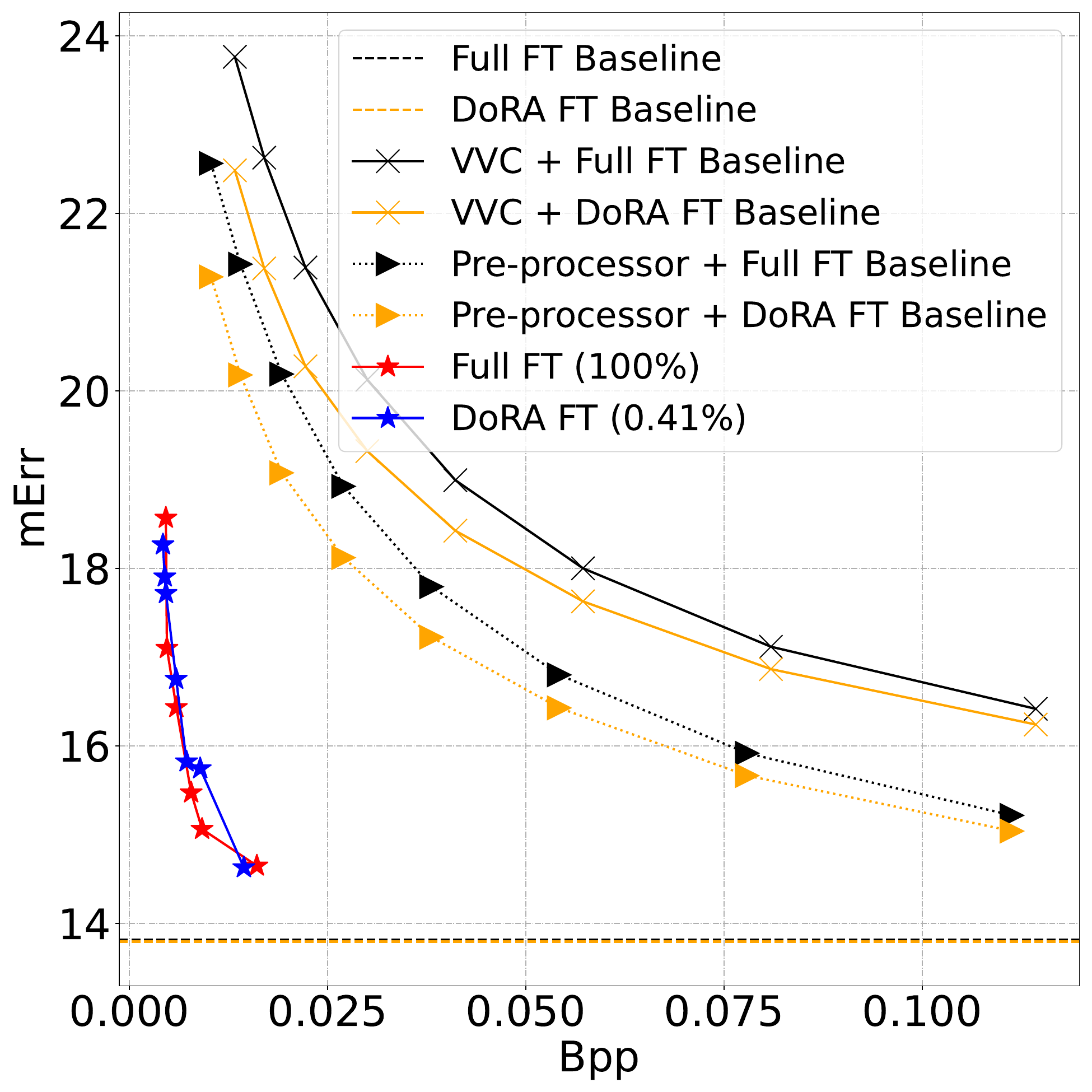}}
\subfigure[Boundary]{\includegraphics[width=\mywidth\linewidth]{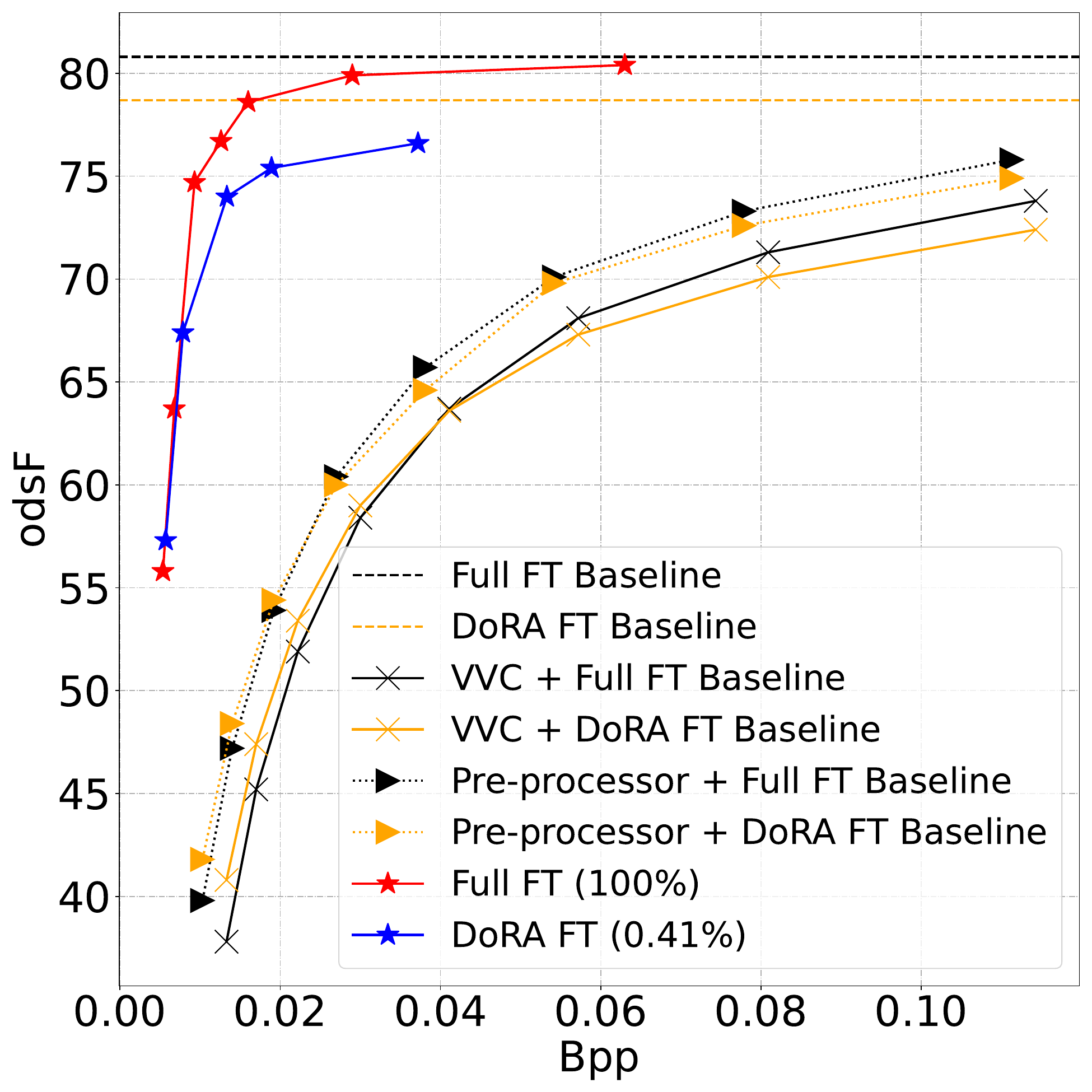}}
\caption{\textbf{RA curves of various methods on PASCAL-Context dataset. }{\it Please zoom in for more details}.} 
\label{fig:pascal_result} 
\end{figure*}

\subsection{Quantitative Results}

We compare our method with the advanced codec VVC~\cite{bross2021overview} and a pre-processor~\cite{lu2024preprocessing}\footnote{As the implementation is not open-sourced and uses different tasks and datasets than ours, we reproduce it based on the description in the paper and enlarge the pre-processor network to enhance performance}, utilizing VVC as the base codec. Additionally, we compare the proposed DoRA fine-tuning (FT) method with the Full FT method.

For the evaluation of VVC, the reconstructed images are fed into the baseline model, which was previously trained on uncompressed images, to generate the corresponding results. The baseline models are identical to those used in the proposed method, except they do not include the compression component. For the evaluation of the pre-processor, images are first processed by the pre-processor model, followed by the VVC evaluation pipeline.

The rate-accuracy curves for all methods are presented in Fig.~\ref{fig:nyud_result} and Fig.~\ref{fig:pascal_result}. In these figures: ``Full FT Baseline'' denotes the baseline model fully fine-tuned without compression, while ``DoRA FT Baseline'' represents the baseline model fine-tuned using DoRA layers without compression. ``VVC + Full FT Baseline'' refers to compressing the image using VVC and then feeding it into the Full FT Baseline model, and similarly, ``VVC + DoRA FT Baseline'' indicates using the DoRA FT Baseline model after VVC compression. ``Pre-processor + Full FT Baseline'' involves processing images through the pre-processor module, compressing them with VVC, and feeding them into the Full FT Baseline model; the same applies to ``Pre-processor + DoRA FT Baseline''. ``Full FT'' represents the fully fine-tuned version of the proposed model, including compression, while ``DoRA FT'' is the final proposed method, which fine-tunes the model using DoRA layers and integrates compression directly within the model.

We summarize the objective criteria and complexities of different methods in Tables 2, 3, and 4 in the supplementary materials. The results from the figures and tables demonstrate that our method significantly outperforms other approaches on dense prediction tasks while requiring substantially less training overhead and resources compared to full fine-tuning or training from scratch. Furthermore, our DoRA fine-tuning achieves performance that is comparable to, or even better than, full fine-tuning. It is important to note that full fine-tuning necessitates storing and deploying a separate copy of the entire backbone for each task, which significantly increases training and storage overhead, thereby limiting its practical applicability. In contrast, our method only requires 0.41\% of the backbone's trainable parameters, compared to 100\% for full fine-tuning, demonstrating the parameter efficiency of our approach. Additionally, the storage required for DoRA layers is minimal, ensuring that our method does not impose a significant storage burden, especially for multi-task scenarios.

\subsection{Ablation Study}

We conduct all ablation experiments on the semantic segmentation task using the NYUD-v2 dataset.
 
\begin{figure}[h] 
\newcommand{\mywidth}{0.49}
\centering 
\subfigure[Pre-trained backbone]{\includegraphics[width=\mywidth\linewidth]{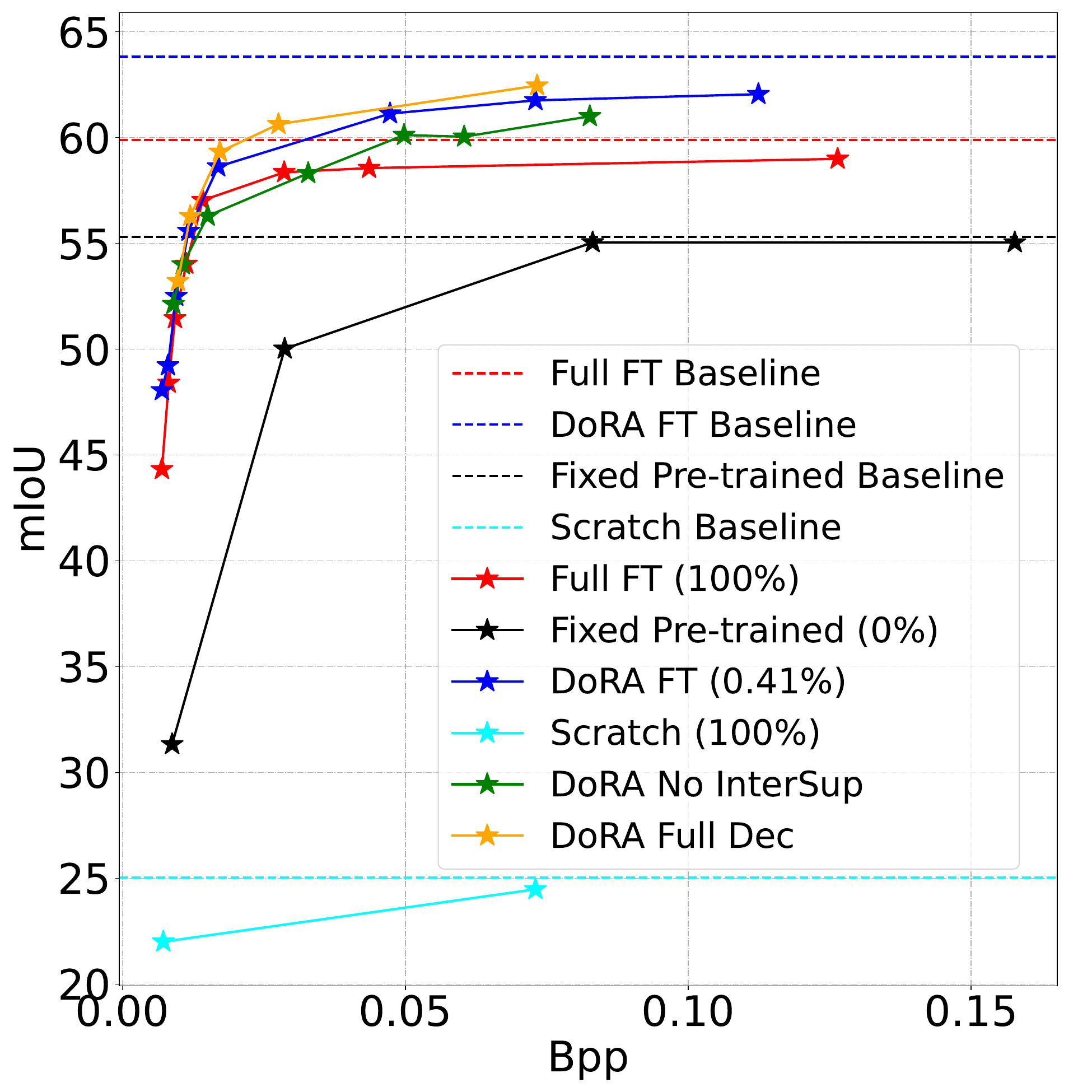}\label{subfig:ab_set}}
\subfigure[DoRA]{\includegraphics[width=\mywidth\linewidth]{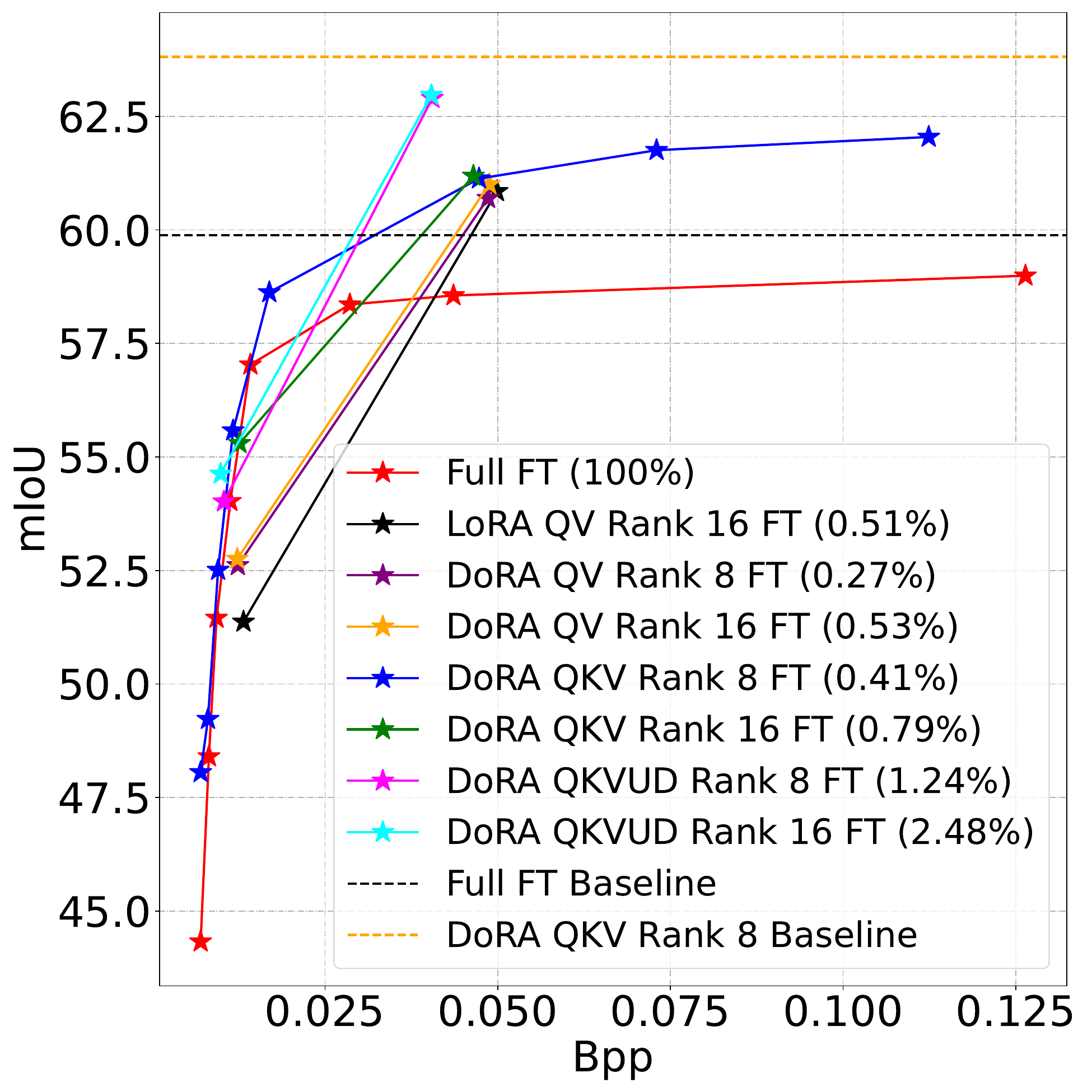}\label{subfig:ab_dora}}
\caption{Ablation study on different settings.}
\vspace{-0.3cm}
\label{fig:ab}
\end{figure}

\textbf{Study on Pre-trained Backbone:}  
First, we investigate the impact of various configurations of the pre-trained vision backbone. The settings evaluated are as follows:  
(1) Full FT: All backbone parameters are fine-tuned.  
(2) DoRA FT: Task-specific DoRA layers are injected, and only these layers are trained.  
(3) Fixed Pre-trained: The pre-trained vision backbone is used directly without fine-tuning.  
(4) Scratch: The backbone is trained from scratch without any pre-training. As shown in Fig.~\ref{subfig:ab_set}, our experiments demonstrate that DoRA FT outperforms Full FT for semantic segmentation. This can be attributed to the limited dataset size, which hinders the convergence of training when all parameters are tuned. Furthermore, we observe that using the Fixed Pre-trained backbone still achieves strong performance, highlighting the robust feature representations of the pre-trained model. In contrast, the Scratch setting performs the worst among all configurations, significantly lagging behind pre-trained models. This highlights the substantial additional training iterations and data required to achieve comparable performance without pre-training. These results emphasize the effectiveness and efficiency of the utilization of the pre-trained vision backbone and DoRAs and validate the strength of our proposed method.

\textbf{Study on Neural Network Architectures:}  
To validate the design of the neural architectures, we conduct the following ablation. First, we evaluate the effectiveness of intermediate task loss supervision in the preliminary decoders. As shown in Fig.~\ref{subfig:ab_set} ``DoRA No InterSup'', eliminating intermediate supervision significantly decreases accuracy, particularly at higher bpp, highlighting the effectiveness of transmitting coarse predictions. Second, we assess the impact of the low-complexity decoder. While employing a full-resolution four stage decoder ``DoRA Full Dec'' provides performance improvement, it incurs substantially higher computational complexity due to the quadratic increase in self-attention with spatial resolution. To achieve a balance between computational efficiency and performance, we adopt the proposed low-complexity decoder as the final design.

\textbf{Study on DoRA Layers:}
Fig.~\ref{subfig:ab_dora} presents an ablation study on the DoRA layers. The percentage indicates the fraction of backbone trainable parameters relative to the backbone's total parameters. We observe that a LoRA layer with a rank of 16, applied to the query and value projections, performs worse than a DoRA layer with ranks of 8 and 16 under comparable trainable parameter settings. Consequently, we select DoRA as the low-rank adaptation method. We further evaluate DoRA applied to the query, key, and value projections with ranks of 8 and 16, as well as to the query, key, value, up, and down projections with the same ranks. The results demonstrate that incorporating more layers and higher ranks generally improves performance due to the increase in trainable parameters. However, to strike a balance between trainable parameters, performance, and energy efficiency, we ultimately select DoRA applied to the query, key, and value projections with a rank of 8 as the final configuration.

\section{Conclusion}
In this work, we propose an energy-efficient framework for ICM that overcomes the limitations of existing methods by leveraging pre-trained vision backbones and task-specific low-rank adaptation layers. Our approach utilizes fixed pre-trained backbones to extract robust and versatile latent representations while refining these features with low-rank adaptation, enabling task-specific optimization with minimal trainable parameters and training energy cost. This design significantly reduces training resources, storage requirements, and computational overhead while maintaining high coding efficiency. Extensive experiments on dense prediction tasks demonstrate that our framework outperforms traditional codecs and pre-processors, and offers superior parameter efficiency and task performance compared to full fine-tuning.


\bibliographystyle{IEEEbib}
{
\footnotesize
\bibliography{multi}
}

\clearpage
\fi
\end{document}